# Persistence of Technosignatures: A Comment on Lingam and Loeb

Milan M. Ćirković[1]
Branislav Vukotić
Milan Stojanović
*Astronomical Observatory of Belgrade, Volgina 7,*
*11000 Belgrade, Serbia*

**Abstract**. In a recent paper in this journal, Lingam and Loeb (2018) develop an excellent heuristic for searches for biosignatures vs. technosignatures. We consider two ways in which their approach could be extended and sharpened, with focus on durability of technosignatures. We also note an important consequence of the adopted heuristic which offers strong support to the ideas of the Dysonian SETI.

There has been a dramatic surge of interest for observing both classical biosignatures and technosignatures of extraterrestrial intelligent species in recent years (e.g., Schneider et al. 2010; Schwieterman et al. 2018). Especially with regard to technosignatures, this development has been a vindication of a number of philosophical and theoretical arguments for more work in this area for quite some time before finding resonance in the mainstream SETI community (Kardashev 1985; Hanson 2008; Bradbury, Ćirković, and Dvorsky 2011; Wright et al. 2014).

In a recent excellent study in this journal, Lingam and Loeb (2018; hence LL18) have ingeniously managed to remove several layers of uncertainty by asking for a *ratio* of the likelihoods of two kinds of astrobiological signatures, instead of trying to calculate their absolute values. This smart move is based on the common evolutionary origin of both conventionally understood biosignatures and technosignatures (including intentional messages and material artefacts), thus showing again how deeply intertwinned are concerns of SETI with those of wider astrobiological and life-sciences fields. In this note, we wish to point out that the kind of reasoning employed by LL18 could be furthered to argue for one particular sector of the SETI studies, namely the search for persistent material artefacts, notably products of astroengineering.

---

[1] Corresponding author; e-mail: mcirkovic@aob.rs. Also at Future of Humanity Institute, Faculty of Philosophy, University of Oxford, Suite 8, Littlegate House, 16/17 St Ebbe's Street, Oxford, OX1 1PT, UK.

The relevant relative likelihood of technosignatures to biosignatures in the case of colonizing civilizations, is given by eq. (11) of LL18 as:

$$\delta \sim 0.02 \left(\frac{f_i}{0.1}\right)\left(\frac{L}{10^4 \text{ yr}}\right)\left[1+\left(\frac{L}{10^4 \text{ yr}}\right)\right], \qquad (1)$$

where the Drake equation terms have their standard meaning ($f_i$ = fraction of biospheres evolving intelligence and technological civilization, $L$ = lifetime of the target civilization). Apart from showing how it follows from the linear colonization model (the rate of "outpost" creation being $\alpha = 1 + L\Lambda$), LL18 do not enter into more detailed discussion of this equation. It is something of an omission, since it is a beautiful and instructive relation.

First, the model of linear colonization growth is likely an underestimate of the real number of sources associated with a higher Kardashev Type (2.x, say) civilization. It is adequate for small-to-moderate scale colonization, which proceeds from a single source. However, it is reasonable to assume that at least some "outposts" become sources of secondary, tertiary, etc. waves of colonization, in which case the total number of technosignatures rises as a higher power of the elapsed time, up to the limiting case of an exponential growth. Even in the linear case, detectability can rise faster under a range of plausible conditions. Suppose that the colonizing efforts created $N$ sites in total, and that the technosignatures under consideration are antimatter-burning engines of interstellar spacecraft travelling between these locations (as per pioneer studies of Harris 1986, 2002); in general case, the number of potentially detectable "routes" will scale $\propto N^2$. Hence, even if colonization of these $N$ sites itself proceeds linearly, the relative likelihood $\delta$ will scale like $L^3$.

One can go even further in building plausible scaling models. If we suppose that the settlement wave moving outward with the average velocity $v$ lasts for the time equal to or proportional to $L$, we can imagine three regimes of settlement. In the first, if $L \ll D_h/v$ ($D_h$ being the vertical scale length of the Milky Way disk), the settlement wave is roughly spherical and the number of sites grows as $L^3$. In the second, when $D_h/v < L < D_r/v$ ($D_r$ being the radial disk scale length), the wave has an axial symmetry and the number of sites grows as $L^2$. Finally, for $L \gg D_r/v$ we have the constant number of sites, since the entire Galaxy is colonized (Hanson 2008). Hence, it is conceivable that the relative likelihood $\delta$ intermittently scales even like $L^4$. Since this is analogous to panspermia – as a kind of "natural settlement" – a similar conclusion has been reached by Ginsburg, Lingam, and Loeb (2018, see in particular their eq. 11).

However, even in the most conservative model of LL18, $\delta$ scales as $L^2$, which has extraordinary long-ranging implications. If civilizations ever approach Kardashev's 2.x Type, a small variation in civilization's age is bound to completely erase the starting bias in favour of biosignatures and shift the balance towards technosignatures. Note also that the normalizing fiducial value for *L* in (1) of $10^4$ years is quite small not only in comparison to either evolutionary or astrophysical timescales, but is also likely to be *small in comparison to the intrinsic scatter* in the distribution of ages of civilizations. Hence the multiplication constant of 0.02 is insufficient to compensate for a likely shift by a factor of a few or an order of magnitude in *L*, and we actually should not conclude on this basis that biosignatures should be favored over technosignatures at all.[*]

Even more important, however, is the fact that the same simple linear model can be used to describe the likelihood of search for the class of astroengineering technosignatures represented by persistent artefacts, such as the Dyson (1960) spheres or artificially shaped occulters (Arnold 2005). Consider first *past human* civilizations on Earth, which we know about mainly on the basis of their artefacts. The ancient Egyptian civilization lasted from the First Dynasty (about 3100 BC; Dee et al. 2013) to the Macedonian conquest in 332 BC; some of their artefacts, like the Great Pyramid of Giza, have already lasted more than that (being finished cca. 2560 BC; cf. Collins 2001). Consider the following durability parameter describing technosignatures:

$$\xi \equiv \frac{\tau_{ts}}{L} \equiv \frac{\text{duration of a particular technosignature}}{\text{the lifetime of its parent civilization}}. \qquad (2)$$

At present, $\xi$ (The Great Pyramid) = 1.65. Obviously, it is increasing insofar as the Great Pyramid continues to resist erosion, weathering, warfare, and other natural and anthropogenic hazards. In some cases, even for human civilizations, the ratio is even higher: $\xi$ (The Pyramid of the Sun in Teotihuacán) = 2.4; $\xi$ (the walls of Cuzco/Sacsayhuamán) is somewhere between 1.8 and 2.5, depending on the unknown exact construction dates, and $\xi$ (Stonehenge) is probably greater than 2.5. Of course, values like these should be taken cautiously, since lifetimes of cultures are hardly ever sharply defined and the construction timescales can be exceedingly long and should be discounted (an analog to this would be the timescale for the Dyson sphere construction, accompanied by its continuously increasing detectability over interstellar distances). However, there is no obvious *upper bound* for $\xi$; it is quite conceivable that the Great Pyramid or Stonehenge would have survived even possible extinction of humanity due to climate change or biological warfare/terrorism. Thus, $\xi$ will continue to rise when most durable artefacts are concerned – and that is true even in the planetary environment, where durability of any structure is substantially reduced by various forms of erosion.

At present, we do not have any way of giving a meaningful value of $\xi$ for artefacts in the interplanetary space, provided that they are in dynamically stable locations (e.g., some of the star-planet Lagrange points) and especially those large enough to be virtually immune to collisions with dust grains, micrometeorites, and even asteroids or comets (though not necessarily functional, cf. DeBiase 2008). There is no reason, however, to presume that it cannot reach large values, from tens to thousands. A civilization might need $10^4$ years to construct a Dyson sphere, and it could become extinct in several times $10^6$ years, while the Sphere itself will continue to exist at least to the end of stellar evolution, which for Sun-like stars occurs on timescales of $10^9$ years or more. Such technosignatures would have $\xi \sim 1000$ or more, making them by far most detectable items on any inclusive list of search targets.

---

[*] One may be tempted to think about biosignatures as persistent as well: after all, if life existed on Mars in the past, its fossils would eventually be found; on the other hand, we may be missing technosignatures of past eras in Earth's history either (Schmidt and Frank 2019). In our view, this points to a kind of "detectability gap" between what is feasible in principle and what is realistic over interstellar – or even intergalactic – distances. In any case, a highly commendable feature of the LL18 study is exactly that it is explicitly pragmatic as far as methodology is concerned. (We thank an anonymous reviewer for bringing our attention to this point.)

Therefore, if there are any civilizations of Kardashev's Type 2.x capable of producing large and stable artefacts, we should invest in searching for them irrespectively of our assumptions about *L*. Even extinct civilizations will leave detectable remnants, which our search for persistent technosignatures might, at least in principle, detect. There is an important point of discontinuity there: while detectability of a civilization increases as it ages and advances in control over physical environment, if it ever reaches the point of decline and fall its non-persistent technosignatures, like directed emissions or leakages, stop rather quickly. This obvious conjecture has in fact been the traditional justification for keeping *L* in all forms of the Drake equation. In contrast, persistent technosignatures, like the astroengineering artefacts, are boosted in the viable target set by a factor of $\xi$ over and above that given by (1). Of course, this represents the point of strongest break with the historical SETI practices, notably search for transient technosignatures like radio-messages. We cannot hope to detect intentional messages, leakages, and callsigns of extinct societies; this probably applies to their beacons as well, which need to be actively powered and maintained structures. As such, even if detectable in principle, we expect beacons to have $\xi$ values *smaller than unity*.

Finally, the square dependence in (1) has the potential to offset all but the most extreme low values of $f_i$. The balance of these two parameters substantially determines the optimal course of our SETI efforts. Let us look at these two from the epistemological point of view. Both „optimists" and „pessimists" agree that $f_i$ is an evolutionary biological issue, and the dividing line is usually along the convergence/contingency debate. The latter has a very long history (Gould 1989; Conway Morris 2003; Lineweaver 2008); in recent years the pendulum has been swinging more in favor of convergence (e.g., Vermeij 2006; Dryden, Thomson, and White 2008; Powell and Mariscal 2015). It has also been claimed that only future research in astrobiology can effectively resolve the dispute (Chela-Flores 2003; Ćirković and Stojković 2017).

Considering the lack of consent in this debate, it is at least arguable whether we can make bigger progress in understanding $f_i$ or *L* in the near-future term. At least on Earth, $f_i$ pertains to reasonably distant and unobservable past, while *L* is closely connected with our current situation and represents the issue of utmost practical significance. Both are likely to suffer from various forms of historical contingency and biases, and both are likely to be much better understood as new complex numerical simulations of evolutionary processes are developed and run. Arguably, we have much stronger motivation for getting better insight into the range and distribution of *L*, but only the future will tell. In any case, this dependency shows, time and again, how astrobiology and SETI are tightly connected to fields such as forecasting and futures studies.

Throughout this comment, we have fully accepted the tacit premise of LL18 (universally accepted in the current astrobiology/SETI discourse) that biosignatures are clearly and unambiguously distinguishable from technosignatures. One interesting speculative possibility which perhaps deserves more attention – and is anyway necessary for the logical closure of discussions of detectability – is that the distinction between biological and technological might eventually be erased. The two could be fused in a sort of "biological artefacts", such as growing/self-repairing megaengineering structures or the Black Cloud of Lem's novel *The Invincible* (Lem [1964] 1973). This is related to the possibilities offered by postbiological evolution – and what comes beyond it (Ćirković 2018). While it is impossible to give a more precise quantitative form to this merging hypothesis yet, we should keep it in mind in any future detailed study. (We thank an anonymous referee for bringing our attention to this radical possibility.)

All in all, the ingenious heuristics of LL18 actually gives strong support to the search for technosignatures, and in particular those persistent technosignatures with long durabilities. As our SETI observations increase in sensitivity and scope, there are reasons for optimism regarding both quantitative and qualitative insights into the parameter space of intelligent life in its most general cosmic context.